\documentclass[prl,aps,twocolumn,showpacs,amsmath,amssymb]{revtex4}
\usepackage{graphicx}
\usepackage{dcolumn}

\begin{document}

\title{SU(4) and SU(2) Kondo Effects in Carbon Nanotube Quantum Dots}

\author{A. Makarovski,$^{1}$ A. Zhukov,$^{1}$ J. Liu,$^{2}$ and G. Finkelstein$^{1}$}

\affiliation{ Departments of  $^1$Physics and $^2$Chemistry, Duke University, Durham, NC 27708}


\begin{abstract}

We study the SU(4) Kondo effect in carbon nanotube quantum dots, where doubly degenerate orbitals form 4-electron ``shells''. The SU(4) Kondo behavior is investigated for one, two and three electrons in the topmost shell. While the Kondo state of two electrons is quenched by magnetic field, in case of an odd number of electrons two types of SU(2) Kondo effect may survive. Namely, the {\it spin} SU(2) state is realized in the magnetic field parallel to the nanotube (inducing primarily orbital splitting). Application of the perpendicular field (inducing Zeeman splitting) results in the {\it orbital} SU(2) Kondo effect.

\end{abstract}


\maketitle

At low temperatures, a variety of nanoscale Coulomb blockade \cite{QDreview} systems with degenerate ground states exhibit the Kondo effect \cite{Hewson}. This many-body phenomenon has now been observed in semiconductor quantum dots, molecules, carbon nanotubes, and magnetic addatoms on metallic surfaces (see Ref. \cite{revival} for a review). In high quality nanotubes the quantum-mechanical orbitals originating in two electronic subbands are doubly degenerate, forming 4-electron ``shells'' (\cite{Liang2002,Buitelaar2002}, see also Ref. \cite{Alex2006} for additional references). In each shell, the Kondo behavior develops in the valleys with one, two, and three electrons \cite{Liang2002,Babic2004,
SU4PRL}. The Kondo effect with one electron in a shell is expected \cite{Choi2005} to obey the SU(4) symmetry \cite{Zarand2003,LeHur2003}, as studied recently in Ref. \cite{Jarillo-Herrero2005} (see also Ref. \cite{Sasaki2004}). In this paper, we investigate the SU(4) Kondo effect in the 1, 2, and 3-electron valleys in magnetic field.

The nanotubes are grown on a $Si/SiO_2$ substrate by Chemical Vapor Deposition using CO as a feedstock gas \cite{Zheng2002}. This method was verified to produce mostly single-wall nanotubes with diameters of about 2 nm. Cr/Au electrodes separated by 200 nm (Figures 1-3) or 600 nm (Figure 4) are deposited on top of the nanotubes. All the measurements are conducted at temperatures between 1.2 K and 2 K. We choose to work with several small-gap semiconducting nanotubes \cite{Cao2004}, which demonstrate high p-type conductance at negative gate voltages. 
At positive gate voltages, the middle section of the nanotube fills with electrons. The part of the nanotube adjacent to the electrodes stays p-type (``leads''). Therefore, a quantum dot is formed {\it within} a nanotube, defined by p-n and n-p junctions. As a result, Coulomb blockade sets in at low temperatures (Figure 1). 

Figure 1 shows conductance map of a 200 nm - long nanotube quantum dot measured as a function of the source-drain bias $V_{SD}$ and gate voltage $V_{gate}$. The ``Coulomb diamonds'' \cite{QDreview} demonstrate clear 4-electron shell filling. The p-n junction transparency grows with $V_{gate}$, resulting in the enhancement of the Kondo effect in each successive shell. The zero-bias Kondo ridge appears in Coulomb diamonds with 1, 2, and 3 electrons (visible for $V_{gate}> 10$ V).  

\begin{widetext}

\begin{figure}[h]
\includegraphics[width=0.85\columnwidth]{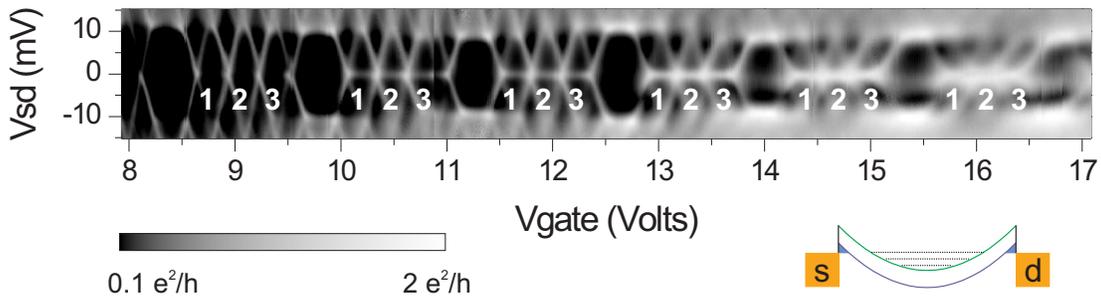}
\caption{\label{fig:intro} 
Differential conductance map of a 200 nm - long semiconducting nanotube quantum dot measured as a function of $V_{gate}$  and $V_{SD}$ (grayscale map: 0.1 to 2 $e^2/h$; $T=2$ K, $B=0$). ``Coulomb diamonds'' demonstrate 4-electron periodicity. Six such 4-electron shells are visible. Contact transparency grows with $V_{gate}$. For $V_{gate}>$ 10 V, the Kondo ridge is visible at $V_{SD}=0$ for 1, 2 and 3 electrons in the topmost shell. Schematic: the quantum dot is formed within the semiconducting nanotube.}
\end{figure}

\end{widetext}

The ambipolar semiconductor nanotubes as studied here are uniquely suited for observation of the SU(4) Kondo effect: the electrons are reflected adiabatically from the p-n junctions at the ends of the quantum dot, resulting in little mode mixing. Therefore, the level mismatch between the two orbitals level in a shell is very small, as evidenced by observation of the Kondo ridge in the 2e valleys of many successive shells (Figure 1). It is also important for the observation of the SU(4) symmetry that the ``leads'' to the dot are formed within {\it the same nanotube}, and thus have the same orbital symmetry, which should be conserved in tunneling processes \cite{Choi2005}.


In nanotubes, parallel magnetic field $B_{\|}$ couples to both the spin and orbital magnetic moment of electrons \cite{Ando2000,Minot2004,Cao2004}. The orbital magnetic moment $\mu$ corresponding to the electron motion around the nanotube circumference is significantly larger than the spin magnetic moment $\mu_0$ (we estimate $\mu \approx 7 \mu_0$, see below). Therefore, in magnetic field the levels in a 4-electron shell should spilt into two doublets moving up or down in energy with $B_{\|}$. Each doublet corresponds to an orbital with a clockwise or counterclockwise direction of rotation with respect to magnetic field, occupied by two electrons (spin-up and spin-down, schematic in Figure 2a). Exactly this behavior is observed in Figure 2a, which shows conductance of a 200 nm-long nanotube quantum dot as a function of the gate voltage and parallel magnetic field. Three 4-electron shells are visible in zero field, which are split into pairs of doublets in $B_{\|}$. 

Figure 2b demonstrates conduction maps measured as a function of $V_{SD}$ and $V_{gate}$ at $B_{\|}=$3 T. The Kondo ridges split at finite field in several horizontal lines at finite $V_{SD}$ (indicated by   triangles) visible inside the Coulomb diamonds. These lines mark inelastic cotunneling thresholds; their appearance indicates that the ground state degeneracy is (partially) lifted. In the cotunneling processes, electrons tunnel in and out of the nanotube through a virtual state, leaving behind an excitation \cite{Averin1992}. The energy of the excitation may be extracted from the value of $e|V_{SD}|$ at which a cotunneling feature is observed. The enhancement of the cotunneling thresholds known as the out-of-equilibrium Kondo effect \cite{Kondo1998,Meir1993,Kogan2004,Zumbuhl2004,Paaske2006} results in the appearance of peaks in the differential conductance (indicated by triangles in Figure 2b).

\begin{figure}[h]
\includegraphics[width=1.0\columnwidth]{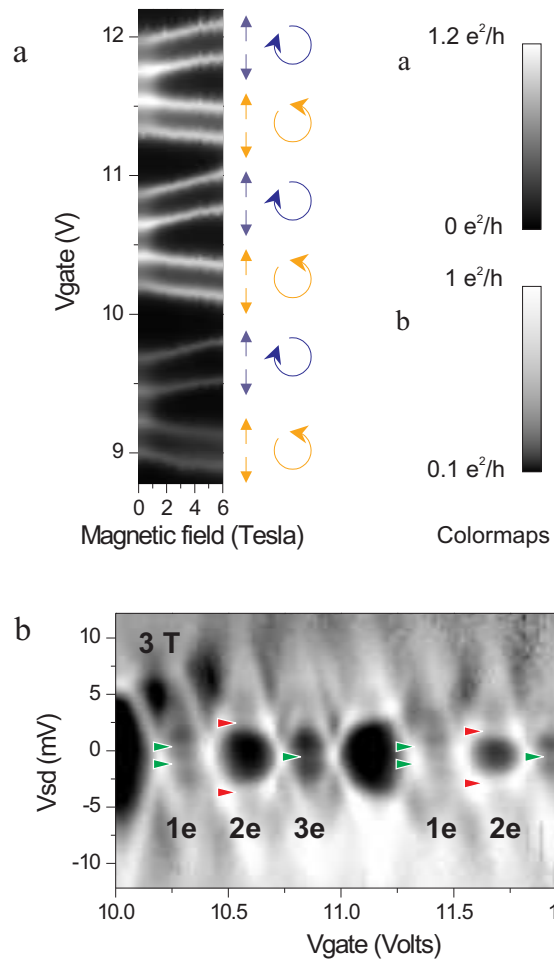}
\caption{\label{fig:orbital} 
a) Differential conductance of a 200 nm - long nanotube quantum dot (similar to Figure 1) measured as a function of $V_{gate}$ and $B_{\|}$. At zero field, three shells are visible. Within each shell, two lower (higher) single-electron traces move down (up) in magnetic field. 
Each doublet corresponds to spin-up and spin-down electrons filling an orbital with a certain direction of rotation in magnetic field. 
Scale of the colormap: 0 to 1.2 $e^2/h$.  
b) Differential conductance map as a function $V_{gate}$ and $V_{SD}$ at $B_{\|}=$ 3 T. Two top shells of Figure 2a are shown. In the 2e valleys, the zero-bias Kondo ridge splits into horizontal cotunneling features at $V_{SD} \approx \pm$ 2 meV (indicated by red triangles), corresponding to an electron excitation from the lower to the higher orbital. The 1e valleys demonstrate two Zeeman-split features at $|V_{SD}| \lesssim$ 1 meV, while the 3e valleys show a single feature close to zero bias (all indicated by green triangles). Grayscale: 0.1 to 1.0 $e^2/h$. }
\end{figure}

The dependence of the co-tunneling features on magnetic field can be best traced in Figure 3, where we show the conductance measured as a function of $V_{SD}$ and $B_{\|}$ in the three valleys. (In each field, the gate voltage is adjusted to stay in the centers of the 1e, 2e, or 3e valleys.) 
Let us consider the 2e valley first. There are six different low-energy states of the two electrons in the nanotube: three different singlet states and one three-component triplet state (schematic in Figure 3b). The energy differences between these states, due to the orbital mismatch, the exchange interaction, and the excess Coulomb interaction, are found to be very small \cite{Alex2006}. In the presence of the lifetime broadening $\Gamma$, the states shown in Figure 3b become effectively degenerate and all participate in the formation of the Kondo resonance. The Kondo state observed here in the two-electron valley is expected \cite{Galpin2005,Martins2007} to obey the SU(4) symmetry. It should be different from the two-electron singlet-triplet Kondo effect induced by level crossing in magnetic field \cite{Sasaki2000}, where only four (and not six) degenerate states participate in the Kondo processes \cite{GPCKreview}.

As $B_{\|}$ is applied, the singlet state with two electrons occupying the lower orbital splits down from the rest (the lowest state in the schematic of Figure 3b). When the energy splitting becomes greater than the Kondo temperature ($\mu B_{\|}\sim k_B T_K$) at $B_{\|} \sim$ 1 T, the zero-bias Kondo ridge disappears (Figure 3b). It is replaced by the inelastic cotunneling thresholds, which correspond to the excitation of the ground state and appear at $eV_{SD} =\pm 2 \mu B_{\|}$ (one of the two electrons is moved from the lower to the higher energy orbital in the shell, see red arrows in the schematic of Figures 3b). These thresholds indeed evolve linearly with field and can be used to estimate the electron angular momentum as $\mu \approx 7 \hbar$ and the diameter of the nanotube as 2 nm \cite{Minot2004,Jarillo-Herrero2005}. 
   
    \begin{widetext}
    
    \begin{figure}[ht]
    \includegraphics[width=0.9\columnwidth]{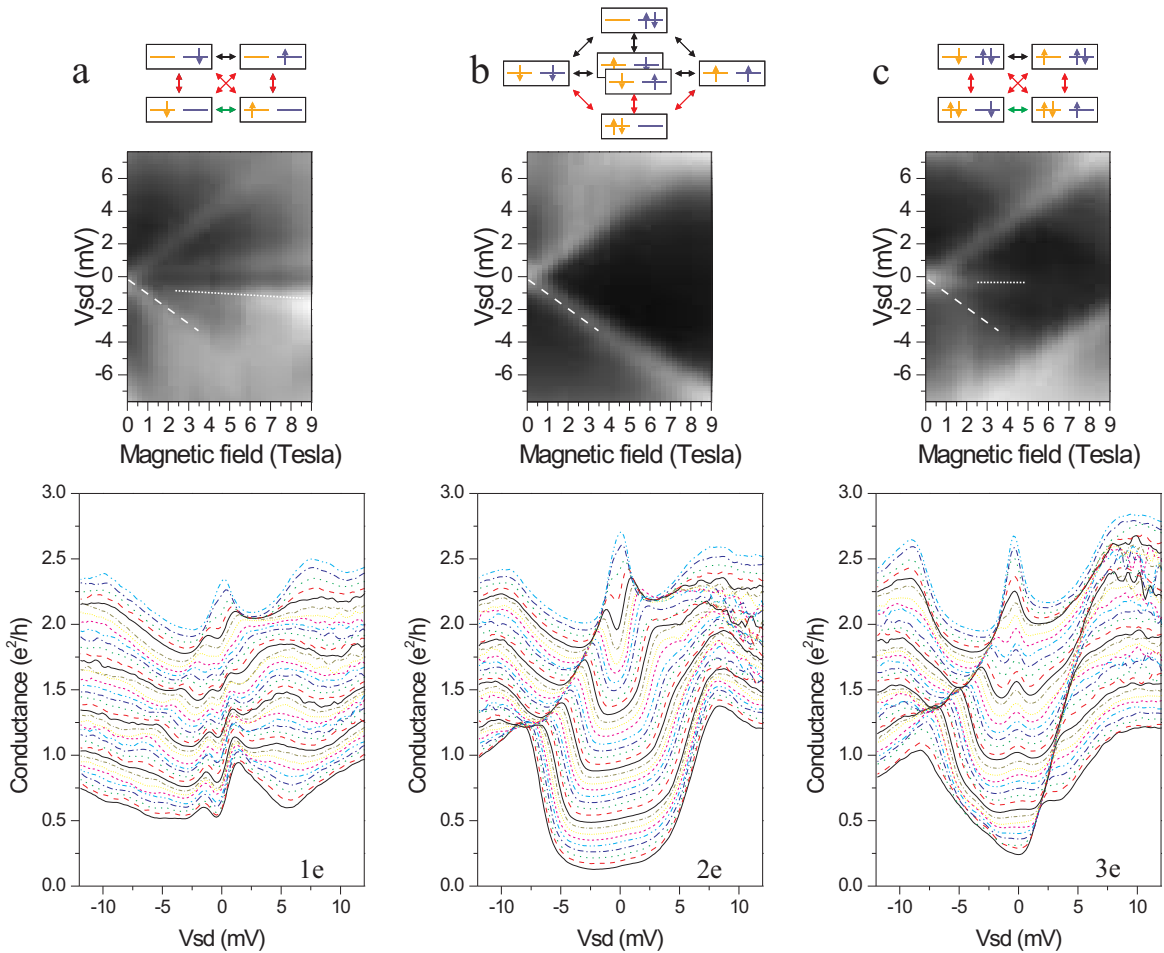}
    \caption{\label{fig:main} 
    Dependence of the a) 1e, b) 2e and c) 3e Kondo features in perpendicular magnetic field. Top: schematics of the different states (rectangular boxes) and the allowed transitions induced between them by the tunneling processes (arrows). These states are degenerate at zero $B_{\|}$, but are split in the field. The lower row represents the states with the lowest orbital energies. The transitions to the states with higher orbital energies are indicated by red arrows. The transitions within the lowest energy Zeeman doublets (for 1e and 3e) are shown by green arrows. Grayscale maps: conductance as a function of the $V_{SD}$ and $B_{\|}$. The gate voltage was adjusted to stay in the center of a valley when the magnetic field was stepped. Kondo zero-bias peak (visible at $V_{SD}$=0, $B_{\|}$=0) splits into four, two, and three features in the 1e, 2e and 3e valleys respectively. The larger energy cotunneling peaks in all three images, marked by dashed lines at negative $V_{SD}$ (and the symmetric features at positive $V_{SD}$), correspond to the orbital splitting. In the 1e valley, the lower energy features marked by a dotted line at negative $V_{SD}$ (and the symmetric feature at positive $V_{SD}$), roughly correspond to the Zeeman splitting. In the 3e valley, the single peak close to zero bias marked by a dotted line survives to higher fields. Lower row: the same differential conductance data shown as a function of the $V_{SD}$ at different $B_{\|}$ ranging from 0 T to 9 T in 0.25 T increments (top to bottom). The curves are offset by 0.05 $e^2/h$ per 0.25 T starting from 9 T. }
    \end{figure}
    
 \end{widetext}
 
In addition to the excitations at $\pm 2 \mu B_{\|}$, the 1e and 3e valleys also demonstrate lower energy features: two resonant cotunneling thresholds at $eV_{SD} \approx \pm g \mu_0 B_{\|}$ in the 1e valley (Figure 3a)  and a zero-bias peak in the 3e valley (Figure 3c). In both valleys, after the orbitals in the shell are split by more than the SU(4) Kondo temperature $T_K^{SU(4)}$, there is one electron left unpaired on the lower or the higher orbital, respectively. This electron can form the SU(2) Kondo state as long as $g \mu_0 B_{\|} \lesssim k_B T_K^{SU(2)}$. Apparently this scenario is realized in the 3e valley, resulting in the appearance of a single Kondo ridge close to zero bias (Figures 2b and 3c). If the SU(2) Kondo temperature is less than $g \mu_0 B_{\|}/k_B$, the SU(2) Kondo state is not formed and cotunneling features at $eV_{SD} \approx \pm g \mu_0 B_{\|}$ should appear, as seen in the 1e valley. A similar 1e behavior was recently observed and attributed \cite{Jarillo-Herrero2005,Choi2005} to the SU(4) Kondo effect.

We observe the difference between the 1e and 3e behaviors in several samples and for several cool-downs. At least in two samples, we can exclude the possibility of the two orbitals in a shell having different $\Gamma$, and hence different $T_K^{SU(2)}$. The effect was also found in successive shells (Figure 2b), so a monotonic change of some parameter with $V_{gate}$ may be ruled out. Overall, while we cannot explain the effect, it appears to be generic. Another open question presented by Figure 3a concerns the low energy cotunneling features. While at high fields their positions $eV_{SD}$ are close to $\pm g \mu_0 B_{\|}$, at low fields they appear at energies significantly exceeding $\pm g \mu_0 B_{\|}$. In particular, the peaks seem to first emerge from the background at the energy of $eV_{SD} \sim k_B T_K^{SU(4)}$ (at $B_{\|} \approx$ 1.5 T). The orbital cotunneling features are already well-formed in this field. We believe these observations call for theoretical interpretations.

We find that the odd-electron valleys also exhibit cotunneling features whose energies {\it decrease} with field. These features are best visible in Figure 3c, in which case the electron occupying the higher orbital of the partially filled shell can be excited to the lower orbital of the next, unoccupied shell. The energy of such an excitation is $\Delta - 2\mu B_{\|}$, where $\Delta$ is the splitting between the shells at zero field. This energy becomes lower in magnetic field, as the corresponding levels come closer. Extrapolating this energy to zero field, we find $\Delta \approx$ 10 meV, consistent with other measurements. Similarly, in 1e valley, an electron can be excited from a lower, completely filled shell, to the lower orbital of the partially filled shell, which is occupied by one electron. This process also requires energy of $\Delta - 2\mu B_{\|}$ (the corresponding faint feature may be visible in Figure 3a). In contrast, in a two-electron valley the inter-shell processes have energy of at least $\Delta$, so that no extra low-energy co-tunneling features are observed.

 \begin{figure}[h]
 \includegraphics[width=0.9\columnwidth]{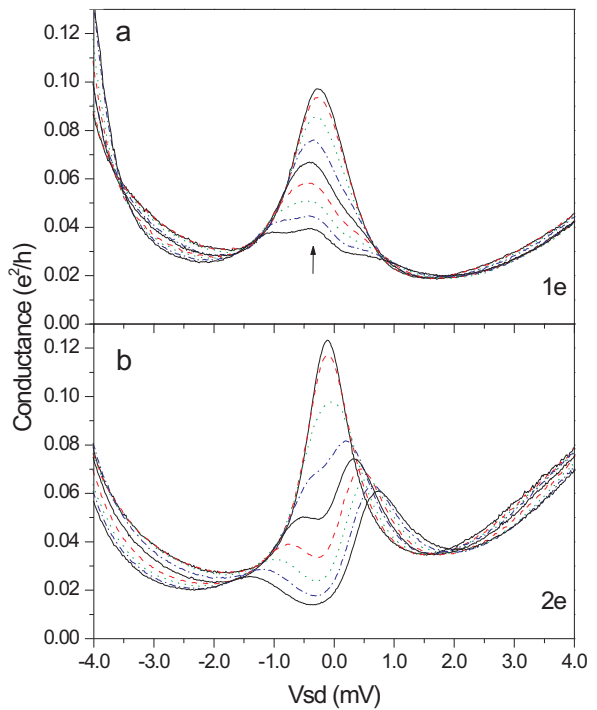}
 \caption{\label{fig:perp} 
 Conductance of a 600 nm-long nanotube quantum dot as a function of $B_{\perp}$ in the a) 1e and b) 2e valleys measured at different magnetic fields. $B_{\perp}$ ranges from 0 T to 8 T (top to bottom) in 1 T increments (the curves are not offset). The vertical arrow indicates the position of the orbital SU(2) peak in the one electron valley.}
\end{figure}

Finally, in Figure 4 we study the dependence of the Kondo conductance on magnetic field perpendicular to the nanotube axis, $B_{\perp}$. In this orientation, the field  primarily couples to the electron spins \cite{Ando2000}. As a result, in the 2e valley, the ground state becomes an electron triplet, with the two electrons occupying different orbitals. The Kondo peak is suppressed (Figure 4b, $B_{\perp} \sim$ 3 T) and non-equilibrium Kondo (cotunneling) features appear at energies slightly above$\pm g \mu_0 B_{\perp}$, corresponding to the spin-flip excitation of one of the electrons. In the 1e valley, on the other hand, the ground state remains degenerate in magnetic field: the electron can occupy one of two orbitals and we may expect to observe the orbital SU(2) Kondo effect \cite{Chudnovskiy2005}. Indeed, the Kondo peak in Figure 4a splits three-ways: the two side peaks correspond to the spin-flip processes, while the center peak corresponds to the orbital SU(2) Kondo effect. The orbital (pseudospin) Kondo effect was observed earlier in double quantum dots \cite{Holleitner2004}.

In conclusion, we study the transitions between the SU(4) and SU(2) Kondo in nanotube quantum dots in magnetic field. The two-electron Kondo effect is suppressed both by parallel and perpendicular magnetic fields, due to formation of a non-degenerate ground state. In parallel magnetic field, the odd-electron SU(4) Kondo may be completely suppressed, or turn into the SU(2) (spin) Kondo effect. In perpendicular magnetic field, the one-electron SU(4) Kondo effect is transformed to SU(2) orbital Kondo effect.

Acknowledgements: We thank H. Baranger, A. Chang, L. Glazman, K. Le Hur, E. Novais, K. Matveev, G. Martins, M. Pustilnik, and D. Ullmo for valuable discussions. The work is supported by NSF DMR-0239748.

\end{document}